\documentstyle[pre,aps,twocolumn,psfig,rotate,amsfonts,eqsecnum]{revtex}
\begin{document}
\def\u{\bbox}
\def\d{\displaystyle}
\def\mathcal#1{{\cal #1}}
\def\goldenmean{\gamma}
\def\phi{\varphi}
\def\epsilon{\varepsilon}
\def\goldenmean{\gamma}

\pagestyle{myheadings}
\markright{Phys. Rev. E {\bf 57}, 1536 (1998). }

\title{Kolmogorov-Arnold-Moser--Renormalization-Group 
Approach to the Breakup of Invariant Tori in Hamiltonian Systems}

\author{C. Chandre, M. Govin, and H. R. Jauslin}

\address{Laboratoire de Physique, CNRS, Universit\'e de Bourgogne,
BP 400, F-21011 Dijon, France}
\maketitle

\begin{abstract}
We analyze the breakup of invariant tori in Hamiltonian
systems with two degrees of freedom using a combination of KAM theory 
and renormalization group techniques.
We consider a class of Hamiltonians quadratic in the
action variables that is invariant under the chosen KAM transformations,
following the approach of Thirring. 
The numerical implementation of the transformation
shows that the KAM iteration converges up to the critical coupling at
which the torus breaks up. By combining this iteration with a renormalization,
consisting of a shift of resonances and rescalings of momentum and energy,
 we obtain a much more 
efficient method that allows to determine the critical coupling with high
accuracy. This transformation is based on the physical mechanism of the
breakup of invariant tori.
We show that the critical surface of the transformation is the
stable manifold of codimension one of a nontrivial fixed point, and we discuss
its universality properties.
\end{abstract}

\section{Introduction}

The existence of invariant tori plays a significant role for the
long-time stability of Hamiltonian systems.
The KAM theorem~\cite{kolmogorov,arnold,moser}
states that the tori with frequency vectors which satisfy 
diophantine conditions are stable under small perturbations.
Conversely, it has been shown~\cite{mackaypercival}
that for large perturbations these tori do no longer exist.
For two-dimensional systems, Greene~\cite{greene,falcolini,mackay3}
formulated a criterion that allows to determine the existence of a KAM 
torus by analyzing the properties 
of a sequence of nearby periodic orbits, namely 
the resonances whose winding ratios are the continued 
fraction approximants of the winding ratio of the considered torus. When the
amplitude of the perturbation is at its critical value, these resonances
open gaps in the torus and it breaks up into a Cantor set 
(Aubry-Mather set)~\cite{aubry,mather,moser2,denzler}.\\
In order to study the self-similar scaling properties observed for the
breakup of invariant tori~\cite{kadanoff,shenkerkadanoff,rand},
renormalization group ideas were proposed for
two-dimensional area-preserving maps~\cite{mackay,mackayL}. 
For Hamiltonian systems with 1.5 degrees
of freedom, Escande and Doveil~\cite{escandedoveil,escande}
set up an approximate renormalization
scheme that combines KAM transformations with a rescaling of phase space.\\
The idea of renormalization group analysis for Hamiltonian systems is  to
construct a transformation $\mathcal{R}$ as a generalized canonical 
change of coordinates acting on some space of Hamiltonians such that the 
iteration of $\mathcal{R}$ converges to a fixed point. If the 
perturbation is smaller than the critical one,
$\mathcal{R}$ must converge to some Hamiltonian
$H_0$ for which the equations
of motion show trivially the existence of a torus of a given frequency 
vector. All Hamiltonians attracted by this trivial fixed point have an
invariant torus
of that frequency: this statement can be considered as an alternative version
of the KAM theorem~\cite{koch}.
If the perturbation is larger than critical, the system does not have a KAM 
torus of the considered frequency and the iteration of $\mathcal{R}$ diverges.
The domain of 
convergence to $H_0$ and the domain of divergence are separated by a surface
invariant under the action of $\mathcal{R}$. The main hypothesis of the 
renormalization group approach is that there should be another nontrivial 
fixed point on this critical surface that is attractive for Hamiltonians 
on that surface. Its existence has strong implications concerning universal 
properties in the mechanism of the breakup of invariant tori.
The analysis of the renormalization for area-preserving maps~\cite{mackay}
gives support
to the validity of this general picture. The aim of the present work
is to give similar support to this picture for Hamiltonian flows.
The main ideas had been announced in Ref.\ \cite{govin}.\\
The transformation we define (KAM-RG) has two main parts: a KAM iteration 
which is a change of coordinates that reduces the size of the
perturbation from $\epsilon$ to $\epsilon^2$, and a 
renormalization transformation which is a combination of a shift of the 
resonances and a rescaling of momentum and energy. It acts within a space of
Hamiltonian systems with two degrees of freedom, quadratic in the action
variables. An essential aspect of the present approach, based on a formulation
of the KAM theorem by Thirring~\cite{thirring}, is that the KAM and 
renormalization transformations we use map this space into itself.
In order to analyze the strong coupling regime and to reach
the critical coupling, the KAM-RG transformation has to converge at least
up to the critical surface at which the torus breaks up. In fact, we show
that the KAM iteration as well as the KAM-RG transformation converge all
the way to the critical coupling. Numerically, the KAM-RG transformation 
is a much more efficient method to determine the critical coupling.
The analysis of the KAM-RG transformation shows that the critical surface is
the stable manifold of codimension one of a nontrivial fixed point.\\
We construct the KAM transformation by two alternative methods: by the Lie
transformations and by transformations defined by a generating function.
The motivation to use two different transformations is two-fold. First we 
verify that both approaches lead to the same results, and the Lie 
transformation is more efficient for numerical implementation. Further,
although the two transformations lead to quantitatively different
nontrivial fixed points, they have the same critical exponents, in
accord with the general ideas of the renormalization group approach.\\
In Sec.\ \ref{sect:ren}, we describe the renormalization transformation.
In Secs.\ \ref{sect:lie} and \ref{sect:tcan}, we define the KAM iteration
of the transformation by the two methods.
In Sec.\ \ref{sect:result}, we give our numerical results, and in particular,
we show evidence of the existence of an even nontrivial fixed point.
In Sec.\ \ref{sect:cat}, we describe the behavior of the KAM-RG transformation 
when odd perturbations are included.\\

We consider the following class of Hamiltonians
with two degrees of freedom,
quadratic in the action variables $\u{A}=(A_1,A_2)$ and described 
by three scalar functions of the angles ${\u \varphi}=(\varphi_1,\varphi_2)$:
\begin{equation}
\label{hamiltonian}
H({\u A},{\u \varphi})=
\frac{1}{2}m({\u \varphi})( {\u \Omega}\cdot{\u A})^{2}
+\lbrack {\u \omega}_{0}
+ g({\u \varphi}){\u \Omega} \rbrack \cdot{\u A}
+ f({\u \varphi}) \ ,
\end{equation}
where ${\u \omega_0}$ 
is the frequency vector of the considered torus, and ${\u \Omega}=(1,\alpha)$ 
is some other constant vector, not parallel to $\u{\omega}_0$. This class
of Hamiltonians has been considered by Thirring~\cite{thirring} in its 
non-degenerate version
\begin{equation}
\label{hamiltonian2}
H({\u A},{\u \varphi})=
\frac{1}{2}\u{A}\cdot M({\u \varphi})\u{A}
+\lbrack {\u \omega}_{0}
+ \u{g}({\u \varphi}) \rbrack \cdot{\u A}
+ f({\u \varphi}) \ ,
\end{equation}
where $M$ is a $2\times2$ matrix such that $\mbox{det} M\not=0$,
and $\u{g}$ a vector.
The Hamiltonian (\ref{hamiltonian}) is such that
\begin{equation}
\mbox{det }\frac{\partial^2 H}{\partial \u{A} \partial \u{A}}=0,
\end{equation}
i.e. it does not satisfy the twist condition, but the KAM theorem is also
valid under this condition (see \cite{gallavotti2}). The advantage of the
family of Hamiltonians~(\ref{hamiltonian}) in the present context is that
they are characterized by three scalar functions of the angles
and a constant $\alpha$, instead
of six functions for Hamiltonians~(\ref{hamiltonian2}). This allows a more
precise numerical treatment of the problem. The essential features are 
already contained in the space of Hamiltonians~(\ref{hamiltonian}). In
particular, the nontrivial fixed point one obtains starting with 
(\ref{hamiltonian2}) is of the form (\ref{hamiltonian}).\\
The functions $m$, $g$, $f$ are represented by their Fourier series e.g.
\begin{equation}
f(\u{\phi})=\sum_{\nu\in {\Bbb Z}^2} f_{\nu} e^{i\nu\cdot\phi}.
\end{equation}
The numerical implementation of the transformation requires a truncation
of the Fourier series. We will approximate $f$ by
\begin{equation}
f^{[\leq L]}(\u{\phi})=\sum_{\nu\in\mathcal{C}_L} f_{\nu} e^{i\nu\cdot\phi},
\end{equation}
where $\mathcal{C}_L=\{\u{\nu}\in {\Bbb Z}^2\left| |\nu_1|\leq L, \, 
|\nu_2|\leq L\right.\}$.
We define $\langle f \rangle$ the mean value of $f$ by
\begin{equation}
\langle f \rangle=\int_{ {\Bbb T}^2}\frac{d^2\u{\phi}}{(2\pi)^2} f(\u{\phi}),
\end{equation}
where ${\Bbb T}^2=[0,2\pi]\times[0,2\pi]$. In the following sections, 
we will use the notation $\u{\partial}f=\d \frac{\partial f}{\partial
\u{\varphi}}$ for any function of the angles.


\section{Renormalization transformation}
\label{sect:ren}

We construct the KAM-RG transformation combining two parts: 
a KAM transformation $(m,g,f,\alpha)\mapsto(m',g',f',\alpha)$ 
and a renormalization consisting of a shift of the resonances and a rescaling
of the actions and of time $(m',g',f',\alpha)\mapsto(m'',g'',f'',\alpha')$.
The renormalization scheme described in this section
is for a torus of frequency $\u{\omega}_0=(1/\goldenmean,-1)$ where
$\goldenmean=(1+\sqrt{5})/2$, but this scheme can be adapted to quadratic
irrationals.\\
The KAM-RG transformation is composed of four steps:\\
1) a KAM transformation which is a change of coordinates that eliminates
terms of order $\mathcal{O}(\epsilon)$, where $\epsilon$ is the size of the
perturbation; this transformation produces terms of the order 
$\mathcal{O}(\epsilon^2)$ and does not change $\u{\Omega}=(1,\alpha)$
(see Secs.\ \ref{sect:lie} and \ref{sect:tcan}),\\
2) a shift of the resonances: a canonical change of coordinates that maps 
the next pair of  daughter resonances of the sequence of rational approximants
into the two main resonances,\\
3) a rescaling of energy (or equivalently of time),\\
4) a rescaling of the action variables (which is
	      a generalized canonical transformation).\\
The aim of this transformation is to treat one scale at the time. 
The steps 2), 3) and 4) are implemented as follows:
The two main resonances $(1,0)$ and $(1,1)$ are replaced by the 
next pair of daughter resonances $(2,1)$ and $(3,2)$, i.e. we require
that $\cos[(2,1)\cdot\u{\phi}']=\cos[(1,0)\cdot\u{\phi}'']$ and 
$\cos[(3,2)\cdot\u{\phi}']=\cos[(1,1)\cdot\u{\phi}'']$.
This change is done via a canonical transformation 
$(\u{A}',\u{\varphi}')\mapsto (N^{-2}\u{A}',N^2\u{\varphi}')$ with 
$$
N^2=\d\left(\begin{array}{cc} 2 & 1 \\ 1 & 1 \end{array}\right).$$
This linear transformation multiplies ${\u \omega}_0$ by $\goldenmean^{-2}$
(since $\u{\omega}_0$ is an eigenvector of $N$);
therefore we rescale the energy by a factor $\goldenmean^2$ in order to keep
the frequency fixed at ${\u \omega}_0$. A consequence of the shift of
the resonances is that $\u{\Omega}$ is changed into $\u{\Omega}'=(1,\alpha')$,
where $\alpha'=(\alpha+1)/(\alpha+2)$.\\
Then we perform a rescaling of the action variables: we change the  
Hamiltonian $H'$ into $$\hat{H'}({\u A}',{\u \varphi}')=
\lambda H'\left(\d\frac{{\u A}'}{\lambda},{\u \varphi}'\right)$$ with $\lambda$
such that the mean value $\langle m'' \rangle$ is equal to $1$.
Since the rescaling of energy and the shift $N^2$ transform the 
quadratic term of the Hamiltonian into 
$\goldenmean^2(2+\alpha)^2m'(\u{\phi}')(\u{\Omega}'\cdot \u{A}')^2/2$, this condition leads to
$\lambda=\goldenmean^2(2+\alpha)^2\langle m' \rangle$.
This condition has the following geometric interpretation in terms of 
self-similarity of the resonances close to the invariant torus:
The rescaling magnifies the size of the daughter resonances,
and places them approximately at the location of the original main resonances.
This can be seen by the following heuristic argument:
In order to estimate the position of the resonances, 
we assume that $\epsilon$ is small and
that $m'\approx\langle m' \rangle.$ The equations of motion for $H'$ are
\begin{eqnarray}
 && \u{A}'\approx const,\\
 && \dot{\u{\phi}'}\approx\langle m' \rangle(\u{\Omega}\cdot\u{A}') \u{\Omega}
 +\u{\omega}_0.
\end{eqnarray}
The position of the resonance $\u{\nu}$ is given by the condition 
$\u{\nu}\cdot\dot{\u{\phi}'}=0$, i.e. it is located at $\u{A}'$ such that
\begin{equation}
\u{\Omega}\cdot\u{A}' \approx -\frac{1}{\langle m' \rangle}
\frac{\u{\omega}_0\cdot\u{\nu}}{\u{\Omega}\cdot\u{\nu}}.
\end{equation}
The linear change of coordinates $N^2$ gives the new position of the
resonance $\u{\nu}$
\begin{equation}
\u{\Omega}'\cdot\u{A}' \approx -\frac{1}{\langle m' \rangle
\goldenmean^2(2+\alpha)^2}
\frac{\u{\omega}_0\cdot\u{\nu}}{\u{\Omega}'\cdot\u{\nu}}.
\end{equation}
Thus, the rescaling of the actions $\u{A}'\mapsto \u{A}''=\u{A}'/\lambda$
with $\lambda=
\goldenmean^2(2+\alpha)^2\langle m' \rangle$ places the resonances at the 
location of the original resonances.\\
In summary, the renormalization rescales $m$, $g$, $f$ and $\u{\Omega}=(1,\alpha)$ into
\begin{eqnarray}
     && m''(\u{\phi})=\d \frac{m'\left(N^{-2}\u{\phi}\right)}
                     {\langle m' \rangle},\\
     && g''(\u{\phi})=\goldenmean^2(2+\alpha) g'\left(N^{-2}\u{\phi}
                      \right),\\
     && f''(\u{\phi})=\goldenmean^4(2+\alpha)^2 \langle m' \rangle
                     f'\left(N^{-2}\u{\phi}\right),\\
     && \alpha'=\frac{1+\alpha}{2+\alpha}.\label{eqn:alpha}
\end{eqnarray}
The iteration of the transformation (\ref{eqn:alpha}) converges to 
$\alpha_*=\goldenmean^{-1}$. It means that ${\u \Omega}$ converges under 
successive iterations  to ${\u \Omega}_*=(1,1/\goldenmean)$, which
 is orthogonal to ${\u \omega}_0$ and is the unstable 
eigenvector of $N^2$ with the largest eigenvalue $\goldenmean^2$.\\
We remark that this renormalization scheme can also be implemented on a 
more general class of Hamiltonians quadratic in the action variables 
considered by Thirring~\cite{thirring}:
\begin{equation}
\label{eqn:hamquadra}
H({\u A},{\u \varphi})=
\frac{1}{2}\u{A}\cdot M({\u \varphi})\u{A}
+\lbrack {\u \omega}_{0}
+ \u{g}({\u \varphi}) \rbrack \cdot{\u A}
+ f({\u \varphi}) \ ,
\end{equation}
where $M$ is a $2\times2$ matrix and $\u{g}$ a vector. This class of
Hamiltonians is also invariant under KAM transformations (see Secs.\
\ref{sect:lie} and \ref{sect:tcan}). The renormalization described above
changes the direction of $\u{g}$ and $M$. 
The vector $\u{g}$ is renormalized into
$\mathcal{R}(\u{g})=\goldenmean^2 N^2\u{g}$. The iteration of $\mathcal{R}$
converges to the unstable eigenvector of $N^2$: $\u{g}\rightarrow g\u{\Omega}_*$.
The matrix $M$ is renormalized into $\mathcal{R}(M)=N^2 M N^2/(N^2 M N^2)_{11}$.
This transformation has only one stable fixed point $M_*=\u{\Omega}_*\otimes
\u{\Omega}_*$. Thus the iteration of the renormalization transformation
on Hamiltonians of the form (\ref{eqn:hamquadra}) (which can satisfy the 
twist condition or not) converges to a twistless Hamiltonian (\ref{hamiltonian})
with $\u{\Omega}=\u{\Omega}_*$.

\section{Lie approach to the KAM transformation}
\label{sect:lie}

The Poisson bracket of two functions of $\u{\phi}$ and $\u{A}$ is given by
\begin{equation}
\lbrace f,g\rbrace =\frac{\partial f}{\partial \u{\varphi}}\cdot
\frac{\partial g}{\partial \u{A}}-\frac{\partial f}{\partial \u{A}}
\cdot\frac{\partial g}{\partial \u{\varphi}}.
\end{equation}
We will work with Lie transformations 
$\mathcal{U}_S: (\u{\varphi},\u{A})\mapsto(\u{\varphi}',\u{A}')$
\begin{eqnarray}
&& \u{A}'=e^{-\hat{S}(A,\varphi)}
           \u{A}\equiv\u{A}-\lbrace S,\u{A}\rbrace
           +\frac{1}{2!}\lbrace S,\lbrace S,\u{A}\rbrace\rbrace\ldots\\
&& \u{\varphi}'=e^{-\hat{S}(A,\varphi)}
                 \u{\varphi}\equiv\u{\varphi}-\lbrace S,\u{\varphi}\rbrace
            +\frac{1}{2!}\lbrace S,\lbrace S,\u{\varphi}\rbrace\rbrace\ldots
\end{eqnarray}
generated by functions $S$ linear in the action variables, of the form
\begin{equation}
\label{eqn:S}
S(\u{A},\u{\varphi})=Y(\u{\varphi})\u{\Omega}\cdot\u{A}
+Z(\u{\varphi})+a\u{\Omega}\cdot\u{\varphi}\ ,
\end{equation}
characterized by two scalar functions $Y$, $Z$, and a constant $a$. 
The expression of the Hamiltonian in these new variables is obtained
by the following equation~\cite{deprit,benettin}
\begin{eqnarray}
H'(\u{A}',\u{\varphi}')&=&
e^{+\hat{S}(A,\varphi)}
H(\u{A},\u{\varphi})\mid_{(A',\varphi')} \nonumber\\
&\equiv& H+\lbrace S,H \rbrace
+\frac{1}{2!}\lbrace S,\lbrace S,H\rbrace\rbrace\ldots
\label{exponential}
\end{eqnarray}
A consequence of the linearity of $S$ in $\u{A}$ is that 
the Hamiltonian $H'$ is again quadratic in the actions, and of the form 
\begin{eqnarray}
\label{image}
  H'({\u A}',{\u \varphi}')&=&
                \frac{1}{2}m'({\u \varphi}')( {\u \Omega}\cdot{\u A}')^{2}
		\nonumber \\
                &+&\lbrack {\u \omega}_0
                + g'({\u \varphi}'){\u \Omega} \rbrack \cdot{\u A}'
                + f'({\u \varphi}').
\end{eqnarray}
We notice that $\u{\Omega}$ is unchanged by this transformation.
Following the approach of Thirring~\cite{thirring}, we consider
the scalar functions $g$ and $f$ of order $\mathcal{O}(\epsilon)$, 
and $m$ of order one.
We determine $S$ such that $g'$ and $f'$ are of order $\mathcal{O}(\epsilon^2)$.
The $n$th iteration of this transformation will produce $g$ and $f$ of order
$\mathcal{O}(\epsilon^{2^n})$ and $m$ of order one.
The idea is that $m(\u{\phi})$ does not need to be eliminated. In order
to show the existence of a torus of
frequency $\u{\omega}_0$ that is located at $\u{A}=0$, it suffices
that the iteration reduces the Hamiltonian (\ref{hamiltonian}) into one
with $f=0$, $g=0$ but $m(\u{\phi})\not=0$. This is an immediate
consequence of the equations of motion associated to 
$H=m(\u{\phi})(\u{\Omega}\cdot\u{A})^2/2+\u{\omega}_0\cdot\u{A}.$ 
The fact that $m(\u{\phi})$ does not need to be eliminated is what allows to
work with canonical transformations that are linear in the actions. This
has the important practical advantage that the KAM transformation leaves 
invariant the subspace of Hamiltonians quadratic in the actions
[Eq.\ (\ref{hamiltonian}) or Eq.\ (\ref{hamiltonian2})].\\
 The expressions of $g'$ and $f'$ up to the order $\mathcal{O}(\epsilon^2)$ are
 \begin{eqnarray}
  g'(\u{\varphi}')&=&g(\u{\varphi}')+\u{\omega}_0\cdot\u{\partial} Y\nonumber \\
                  &&+m(\u{\varphi}')(\u{\Omega}\cdot\u{\partial} Z
                    +a\Omega^2) +\mathcal{O}(\epsilon^2),\\
 f'(\u{\varphi}')&=&f(\u{\varphi}')+\u{\omega}_0\cdot\u{\partial} Z
 + a \u{\omega}_0\cdot\u{\Omega}+\mathcal{O}(\epsilon^2).
 \end{eqnarray}
Thus in order to eliminate the terms of order $\mathcal{O}(\epsilon)$,
we determine:\\
1) $Z(\u{\varphi})$ such that the term independent of the action variables
is of the order $\mathcal{O}(\varepsilon^{2})$. The function $Z$ must satisfy the 
equation
\begin{equation}
\label{eqn:lie1}
f+\u{\omega}_0\cdot\u{\partial}Z=const,
\end{equation}
which has the solution
\begin{equation}
Z(\u{\varphi})=\sum_{\nu\not=0} \frac{i}{\u{\omega}_{0}\cdot\u{\nu}}f_{\nu}
e^{i\nu\cdot\phi}.
\end{equation}
The mean value of $Z$ is not determined by Eq.\ (\ref{eqn:lie1}).
We choose it equal to zero.\\
2) Next, we determine $a$ and $Y(\u{\varphi})$ such that the linear 
term in the action variables 
becomes of the form 
$\lbrack \u{\omega}_{0}+\mathcal{O}(\varepsilon^{2})\u{\Omega}
\rbrack \cdot\u{A}'.$ This leads to the condition
\begin{equation}
\label{eqn:lie2}
g+\u{\omega}_0\cdot\u{\partial}Y+m\u{\Omega}\cdot\u{\partial}Z
+m a\Omega^2 =0,
\end{equation}
which has the solution
\begin{equation}
a=-\frac{\langle g\rangle+\langle m\u{\Omega}\cdot\u{\partial}Z\rangle}
{\Omega^2\langle m \rangle},
\end{equation}
and
\begin{equation}
Y(\u{\varphi})=\sum_{\nu\not=0} \frac{i}{\u{\omega}_{0}\cdot\u{\nu}}
\left(g_{\nu}+(m\u{\Omega}\cdot\u{\partial}Z)_{\nu}
+m_{\nu}a \Omega^2\right)e^{i\nu\cdot\phi}\ .
\end{equation}
The transformed Hamiltonian (\ref{image}) is constructed by defining
$H^{(0)}=H$ and $H^{(i)}$ for $i=1,2,\ldots$ by the recursive relation
\begin{eqnarray}
H^{(i+1)}(\u{A},\u{\varphi})&=&
\lbrace S(\u{A},\u{\varphi}),H^{(i)}(\u{A},\u{\varphi}) \rbrace \nonumber\\
&=&
\frac{1}{2}m^{(i+1)}(\u{\varphi})(\u{\Omega}\cdot\u{A})^{2}\nonumber \\
&& +g^{(i+1)}(\u{\varphi})\u{\Omega}\cdot\u{A}+f^{(i+1)}(\u{\varphi}),
\end{eqnarray}
which leads to
\begin{equation}
H'=\sum_{i=0}^\infty \frac{H^{(i)}}{i!}.
\end{equation}
This can be expressed in terms of the image of the three scalar functions 
$(m,g,f)$ given by the following equations:
\begin{eqnarray}
 && (m',g',f')=\left(\sum_{i=0}^{\infty} \frac{m^{(i)}}{i!},
\sum_{i=0}^{\infty} \frac{g^{(i)}}{i!},\sum_{i=0}^{\infty} \frac{f^{(i)}}{i!}
\right),\\
 && (m^{(0)},g^{(0)},f^{(0)})=(m,g,f),
\end{eqnarray}
\begin{eqnarray}
 m^{(1)}&=&2m\u{\Omega}\cdot\u{\partial}Y
           -Y\u{\Omega}\cdot\u{\partial}m,\\
 g^{(1)}&=&g\u{\Omega}\cdot\u{\partial}Y
           -Y\u{\Omega}\cdot\u{\partial}g \nonumber \\
           &&+m\u{\Omega}\cdot\u{\partial}Z+m a\Omega^2
           +\u{\omega}_0\cdot\u{\partial}Y,\\
 f^{(1)}&=&-Y\u{\Omega}\cdot\u{\partial}f
           +g\u{\Omega}\cdot\u{\partial}Z+g a\Omega^2 
           +\u{\omega}_0\cdot\u{\partial} Z, \\
 m^{(i+1)}&=&2m^{(i)}\u{\Omega}\cdot\u{\partial}Y
             -Y\u{\Omega}\cdot\u{\partial}m^{(i)}, \\
 g^{(i+1)}&=&g^{(i)}\u{\Omega}\cdot\u{\partial}Y
             -Y\u{\Omega}\cdot\u{\partial}g^{(i)} \nonumber \\
             &&+m^{(i)}\u{\Omega}\cdot\u{\partial}Z+m^{(i)}a\Omega^2,\\
 f^{(i+1)}&=&-Y\u{\Omega}\cdot\u{\partial}f^{(i)}
             +g^{(i)}\u{\Omega}\cdot\u{\partial}Z+g^{(i)}a\Omega^2,
\end{eqnarray}
for $i\geq 1$.


\section{Generating function approach to the KAM transformation}
\label{sect:tcan}

In this section we describe another construction of the KAM 
transformation. The transformation is now taken as a canonical 
transformation 
$\mathcal{U}_F:(\u{\phi},\u{A})\mapsto (\u{\phi}',\u{A}')$ 
defined by a generating function~\cite{goldstein} characterized by two
scalar functions $Y$, $Z$, of the angles, and a constant $a$, of the form
\begin{equation}
\label{eqn:Scan}
F(\u{A}',\u{\phi})=\left(\u{A}'+a\u{\Omega}\right)\cdot\u{\phi}
+ Y(\u{\phi})\u{\Omega}\cdot\u{A}'+ Z(\u{\phi}),
\end{equation}
leading to
\begin{eqnarray}
&&  \label{eqn:tcan1}
    \u{A}=\d \frac{\partial F}{\partial \u{\phi}}
	 =\u{A}'+\u{\Omega}\cdot\u{A}'\u{\partial}Y
                  + a \u{\Omega} 
                  + \u{\partial} Z,\\
&& \label{eqn:tcan2}
    \u{\phi}'=\d \frac{\partial F}{\partial \u{A}'}
             =\u{\phi}+Y(\u{\phi})\u{\Omega}.
\end{eqnarray}
Inserting Eq.\ (\ref{eqn:tcan1}) into Eq.\ (\ref{hamiltonian}), we obtain the 
expression of the Hamiltonian in the mixed representation
of new action variables and old angle variables
\begin{eqnarray}
\label{hamaction}
\tilde{H}({\u A}',{\u \varphi})&=&
\frac{1}{2}\tilde{m}({\u \varphi})( {\u \Omega}\cdot{\u A}')^{2}\nonumber \\
&&+\lbrack {\u \omega}_{0}
+ \tilde{g}({\u \varphi}){\u \Omega} \rbrack \cdot{\u A}'
+ \tilde{f}({\u \varphi}) \ ,
\end{eqnarray}
with 
\begin{eqnarray}
 && \tilde{m}=(1+\u{\Omega}\cdot\u{\partial}Y)^2 m,\\
 && \tilde{g}=g+\u{\omega}_0\cdot\u{\partial}Y+m b+
            \u{\Omega}\cdot\u{\partial}Y\left(g+m b\right),\\
 && \tilde{f}=f+\u{\omega}_0\cdot\u{\partial}Z+\frac{1}{2}m b^2 +g b,
 \label{eqn:tcanf}
\end{eqnarray}
where $b(\u{\phi})=a\Omega^2+\u{\Omega}\cdot\u{\partial}Z$. We notice that 
the KAM transformation does not change $\u{\Omega}$.\\
We determine the generating function (\ref{eqn:Scan})
such that $\mathcal{H}\circ\mathcal{U}_F$ 
vanishes to the first order in $\epsilon$.
This leads to the conditions
\begin{eqnarray}
   && \label{eqn:cond1}
   \u{\omega}_0\cdot\u{\partial}Z+f=const,\\
   && \label{eqn:cond2}
   \u{\omega}_0\cdot\u{\partial}Y+g
   +m\left(a\Omega^2+\u{\Omega}\cdot\u{\partial} Z\right)=0.
\end{eqnarray}
We recall that the functions $g$ and $f$ are of order 
$\mathcal{O}(\epsilon)$ and $m$ is of order one; 
as a consequence $Y$, $Z$ and $a$ are of order $\mathcal{O}(\epsilon)$.
We notice that these equations are the same as Eqs.\ (\ref{eqn:lie1}) and
(\ref{eqn:lie2}) which determine
the generator of the Lie transformation. The present transformation and the
Lie transformation described in Sec.\ \ref{sect:lie} 
are canonical transformations
with identical linear part [i.e. $\mathcal{O}(\epsilon)$ part]
but different nonlinear terms (of the higher order in $\epsilon$).
The main practical difference
with the Lie transformation is that Eq.\ (\ref{eqn:tcan2}) which determines
the new angles has to be inverted.
Equations (\ref{eqn:cond1}) and (\ref{eqn:cond2}) are solved by representing them 
in Fourier space. They define the generating function $F$ as
\begin{eqnarray}
  && Z(\u{\phi})=\sum_{\nu\not= 0} \frac{i}{\u{\omega}_0\cdot\u{\nu}}
                 f_{\nu} e^{i\nu\cdot\phi},\\
  && a=-\frac{\langle g\rangle+\langle m\u{\Omega}\cdot\u{\partial}Z\rangle}
          {\Omega^2\langle m \rangle},\\
  && Y(\u{\phi})=\sum_{\nu\not= 0} \frac{i}{\u{\omega}_0\cdot\u{\nu}}
                 \left(g_{\nu}+(m\u{\Omega}\cdot\u{\partial}Z)_{\nu} \right.
		 \nonumber \\
  &&	\qquad \qquad \qquad \qquad	 \left. +m_{\nu} a \Omega^2\right)
                  e^{i\nu\cdot\phi}.
\end{eqnarray}
Thus the scalar functions of $\tilde{H}$ become
\begin{eqnarray}
 && \tilde{m}=(1+\u{\Omega}\cdot\u{\partial}Y)^2 m,\\
 && \tilde{g}=-\u{\omega}_0\cdot\u{\partial}Y \u{\Omega}\cdot\u{\partial}Y,\\
 && \tilde{f}=\frac{1}{2}(g-\u{\omega}_0\cdot\u{\partial}Y)
              \left(a\Omega^2+\u{\Omega}\cdot\u{\partial}Z\right).
\end{eqnarray}
We notice that $\tilde{m}$, $\tilde{g}$ and $\tilde{f}$ are given by products 
and sums of functions 
whose Fourier coefficients are explicitly known. We expand these functions in
Fourier series e.g. 
$\tilde{m}(\u{\phi})=\sum_{\nu} \tilde{m}_{\nu}e^{i\nu\cdot\phi}$.\\
The expression of the Hamiltonian
in the new angle variables requires the inversion of Eq.\ (\ref{eqn:tcan2}). 
The Jacobian of this transformation is 
\begin{equation}
\left| \mathrm{det} \left(\d\frac{\partial \phi_k'}{\partial \phi_l}
\right) \right|=\left| 1+\u{\Omega}\cdot\u{\partial}Y\right|.
\end{equation}
The scalar functions $m'$, $g'$ and $f'$ are respectively $\tilde{m}$, 
$\tilde{g}$ and $\tilde{f}$ expressed 
in the new angle variables e.g. $m'(\u{\phi}')=\tilde{m}[\u{\phi}(\u{\phi}')]$.
The Fourier coefficients of $m'$ are determined by the following integrals
\begin{equation}
 \label{eqn:coeff}
 m_{\nu}'=\int_{ {\Bbb T}^2}
 \frac{d^2\u{\phi}'}{(2\pi)^2} m'(\u{\phi}') 
           \, e^{-i\nu\cdot\phi'}.
\end{equation}
With the change of variables 
$\u{\phi}'\mapsto \u{\phi}$, we can write
\begin{equation}
 m_{\nu}'=\int_{ {\Bbb T}^2} 
 \frac{d^2\u{\phi}}{(2\pi)^2} \left|
       \mathrm{det} \left(\d\frac{\partial \phi_k'}{\partial \phi_l}\right) 
       \right| \tilde{m}(\u{\phi}) 
       \, e^{-i\nu\cdot(\phi+Y(\phi)\Omega)}.
\end{equation}
Therefore $m'_{\nu}$, $g'_{\nu}$ and $f'_{\nu}$ can be expressed in the 
following way:
\begin{eqnarray}
  && m'_{\nu}=\sum_{\nu'}\tilde{m}_{\nu'} C_{\nu'\nu}, \quad
  g'_{\nu}=\sum_{\nu'}\tilde{g}_{\nu'} C_{\nu'\nu}, \nonumber \\
  && f'_{\nu}=\sum_{\nu'}\tilde{f}_{\nu'} C_{\nu'\nu},
\end{eqnarray}
where 
\begin{equation}
C_{\nu'\nu}=\int_{ {\Bbb T}^2} 
\frac{d^2\u{\phi}}{(2\pi)^2} \, 
                |1+\u{\Omega}\cdot\u{\partial}Y| e^{i(\nu'-\nu)\cdot\phi}
		e^{-i\nu\cdot\Omega Y(\phi)}. \label{eqn:int}
\end{equation}
In order to compute $C_{\nu'\nu}$,
we choose the Gauss quadrature that approximates an integral
as a sum over a lattice of constant step:
\begin{eqnarray}
&&\int_{ {\Bbb T}^2} 
\frac{d^2\u{\phi}}{(2\pi)^2} \; G(\u{\phi})=\nonumber \\
&&\lim_{M\rightarrow\infty}
\frac{1}{(M+1)^2}\sum_{i,j=0,\ldots,M} G\left(\frac{2\pi}{M+1}i,
\frac{2\pi}{M+1}j\right).
\end{eqnarray}
This quadrature is exact if all the non-zero Fourier modes 
of the considered function $G$ are inside $\mathcal{C}_M$.
For instance in the case of the computation of $C_{\nu'\nu}$ for the 
identity transformation ($Y=0$), since $\u{\nu}'-\u{\nu} \in \mathcal{C}_{2L}$,
 one needs to take $M \geq 2L$. For the general case, if we expand 
the function to be integrated as a power series of $\epsilon$, we notice that
the algorithm with $M=(k+1)L$ gives an approximation up to
$\mathcal{O}(\epsilon^k)$. For instance this algorithm with $k=5$ gives 
an accurate approximation of the exact transformation and allows to compute 
all its properties.
	      

\section{Determination of the critical coupling; fixed point of the KAM-RG 
transformation}
\label{sect:result}

We start with the same initial Hamiltonian as in
Refs.\ \cite{escande,cellettichierchia} 
\begin{equation}
\label{hamiltonieninit}
H({\u A},{\u \varphi})=\frac{1}{2}({\u \Omega}\cdot{\u A})^{2}
+{\u \omega}_{0}\cdot{\u A}+\varepsilon f({\u \varphi}) \ ,
\end{equation}
where ${\u \Omega}=(1,0)$, ${\u \omega}_0
=(1/\goldenmean,-1)$,
$\goldenmean=(1+\sqrt{5})/2$, and a perturbation
\begin{equation}
f({\u \varphi})=\cos({\u \nu}_1\cdot{\u \varphi})
+ \cos({\u \nu}_2\cdot{\u \varphi}), 
\end{equation}
where ${\u \nu}_1=(1,0)$ and ${\u \nu}_2=(1,1)$.
We perform an iteration of the KAM transformation described
in Sec.\ \ref{sect:lie} or in Sec.\ \ref{sect:tcan}. The
two methods give qualitatively the same results. The algorithm using the Lie 
transformation is numerically more efficient. The
following results are those obtained by the Lie transformation.
We represent all the
functions by their Fourier series truncated by retaining only the coefficients in
the square $\mathcal{C}_L$ which contains $(2L+1)^2$ Fourier coefficients. 
For fixed $L$ we take successively larger couplings
$\varepsilon$ and determine whether the KAM iteration converges to a Hamiltonian with 
$f=0,\,  g=0$, or whether it diverges ($f,g\rightarrow\infty$).
By a bisection procedure, we determine the critical
coupling $\varepsilon_c(L)$. We repeat the calculation with larger numbers
of Fourier coefficients, to obtain a more accurate approximation. In Fig.\ 1,
we show $\varepsilon_c(L)$ i.e.  the dependence of the critical coupling on the
number of Fourier coefficients retained. We observe that $\varepsilon_c(L)$ 
decreases with $L$ in a stepwise manner. It stays essentially constant except at the
values of $L$ where a new rational approximant of the frequency ${\u \omega_0}$ is included,
corresponding to a resonance at the next smaller scale. The size of the jumps
diminishes approximately geometrically, and we can extrapolate to obtain the value
$\varepsilon_c(L) \to 0.0276$. This value is close to the
critical coupling $\varepsilon_c=0.0275856$ obtained by the Greene
criterion\cite{greene,cellettichierchia} which is surmised to yield the exact
value. This gives a numerical evidence that the KAM iteration can be expected
to converge in the whole domain of existence of the torus.\\
In Figs.\ 1 and 2, we show the values of 
the critical coupling $\varepsilon_c(L)$,
calculated by the KAM-RG transformation
which is a combination of a Lie transformation (Sec.\ \ref{sect:lie})
and a renormalization transformation (Sec.\ \ref{sect:ren}). We obtain
$\varepsilon_c\in[0.027585,0.027595]$, which is in very good agreement with the
value  $\varepsilon_c=0.0275856$ obtained with the Greene criterion.
We observe that the KAM-RG transformation gives very high precision already with few
Fourier coefficients e.g. $\epsilon_c(L=5)=0.0276633$.\\
The improvement with respect  to the  KAM iteration is not only
quantitative; the disappearance of the steps is a strong evidence that the KAM-RG
transformation we have constructed captures the essential physical
mechanism of the breakup of the tori.\\
By iterating the KAM-RG transformation starting from a point on the critical
surface, we observe that the process converges to a nontrivial critical point
$H_*$,
which we characterize by the Fourier coefficients of the three functions
 $f_*,g_*,m_*$ and ${\u \Omega}_*=(1,\goldenmean^{-1})$.
Figure 3 shows the weight of the Fourier coefficients of $f_*$. We observe that
the nonzero coefficients are strongly concentrated on a band around the direction 
$D_\perp$ perpendicular to the line of resonances.
$D_\perp$ is the expansive direction of the map $\u{\nu}\mapsto
N^{-2}\u{\nu}$, i.e. the direction of the frequency vector $\u{\omega}_0$.
The  decrease of the size of the coefficients along $D_\perp$
is quite slow. The Fourier coefficients of $g_*$ and $m_*$ have a
similar overall behavior, but they decay faster in the $D_\perp$ direction
(see Figs.\ 4 and 5). 
By linearizing the KAM-RG transformation around the fixed point 
$H_*$, we calculate the critical exponents. 
There is only  one with modulus greater than one, denoted by $\delta$. This 
implies that the critical surface, which is the stable manifold 
of $H_*$, is of codimension one.
The value we obtain for the relevant critical exponent is $\delta\in[2.67,2.68]$
which is quite close to the one obtained by MacKay for area-preserving
maps\cite{mackay} ($\delta=2.65$), and to the one obtained by Escande 
{\it et al.}\ with the approximate scheme ($\delta=2.75$)~\cite{escandeetal}.\\
For the scaling factor at the nontrivial fixed point, 
we obtain numerically $\lambda_*=
17.944$, which is very close to $\goldenmean^6$ (the scaling factor at
the trivial fixed point). This value can be compared with the one given for
area-preserving maps $\lambda_*=18.827$ obtained in Refs.\ 
\cite{kadanoff,shenkerkadanoff,mackay}.\\
We remark that if we consider the more general starting Hamiltonian of the
form (\ref{hamiltonian2}) $H=\u{A}\cdot M \u{A}/2 +\u{\omega}_0\cdot \u{A} +f$
with the twist condition $\mbox{det}M\not=0$, the KAM-RG iteration has the
same nontrivial fixed point as we have found for twistless Hamiltonians 
(\ref{hamiltonian}).\\
We have also performed this analysis using the KAM transformation constructed
with a generating function as described in Sec.\ \ref{sect:tcan}. The results
(using fewer Fourier coefficients than the Lie transformation) are qualitatively
similar. The nontrivial fixed point is quantitatively different, but the 
critical exponents are the same. This is what is usually expected in a
renormalization group approach.


\section{Symmetries of the transformation: generalization to 
non-even perturbations}
\label{sect:cat}

In the previous section, we found numerically the existence of a nontrivial fixed 
point for the KAM-RG transformation starting with an even initial perturbation
on the critical surface
\begin{equation}
\label{pert1}
f_0(\u{\phi})=-M\cos(\u{\nu}_1\cdot\u{\phi})-P\cos(\u{\nu}_2\cdot\u{\phi}),
\end{equation}
where $M,P\geq 0$ and $\u{\nu}_1$, $\u{\nu}_2$ correspond to the two
main resonances. For a perturbation containing also odd terms
$\sin(\u{\nu}_1\cdot\u{\phi})$ and $\sin(\u{\nu}_2\cdot\u{\phi})$, 
the KAM-RG transformation, acting on the critical surface, does not necessarily
converge to a fixed point but to a periodic or even a nonperiodic orbit
(we shall see that these attracting orbits are those of the Arnold's cat map).
Already if we start with the even perturbation (\ref{pert1}) with
$M,P\leq 0$, we obtain a cycle of period three as it has also been encountered
in area-preserving maps~\cite{greenemao}. This can be understood by the
symmetries of the transformation~\cite{koch,mackay2}. 
Starting with the two main resonances
$\u{\nu}_1,\u{\nu}_2$, the most general perturbation can be written
as 
\begin{eqnarray}
\label{pert2}
f_0(\u{\phi})&=&-M_e\cos\left(\u{\nu}_1\cdot\u{\phi}\right)
-P_e\cos\left(\u{\nu}_2\cdot\u{\phi}\right)\nonumber\\
&&-M_o\sin\left(\u{\nu}_1\cdot\u{\phi}\right)
-P_o\sin\left(\u{\nu}_2\cdot\u{\phi}\right), \\
&=&-M\cos\left(\u{\nu}_1\cdot(\u{\phi}+\u{\theta})\right)
-P\cos\left(\u{\nu}_2\cdot(\u{\phi}+\u{\theta})\right),\nonumber
\end{eqnarray}
where $M=\sqrt{M_e^2+M_o^2}$, $P=\sqrt{P_e^2+P_o^2}$,\\
and $$\u{\theta}=\left(
-\arctan(M_o/M_e) , \arctan(M_o/M_e)-\arctan(P_o/P_e) \right).$$ 
The question is to analyze the 
effect of a shift of the angles on the transformation. We define a shift
operator by
\begin{equation}
\mathcal{T}_{\theta}:\u{\phi}\mapsto\u{\phi}+\u{\theta}.
\end{equation}
The KAM transformation commutes with $\mathcal{T}_{\theta}$,
as can be easily verified from Eq.\ (\ref{exponential}) or Eqs.\ 
(\ref{eqn:tcan1})-(\ref{eqn:tcanf}). 
The action of 
$\mathcal{T}_{\theta}$ on the shift of the resonances $N^2$ is described 
by the following intertwining relation:
\begin{equation}
\mathcal{R}\circ\mathcal{T}_{\theta}=\mathcal{T}_{N^2\theta}\circ
\mathcal{R},
\end{equation}
where $\mathcal{R}$ denotes the KAM-RG transformation.
Applying this relation to the fixed point $H_*(\u{\phi})$, 
we have $\mathcal{R}H_*(\u{\phi}+\u{\theta})=H_*(\u{\phi}+N^2\u{\theta})$.
The following map (which is Arnold's cat map~\cite{arnoldavez})
\begin{equation}
\label{eqn:cat}
\u{\theta}\mapsto N^2\u{\theta} \mbox{ mod } 2\pi,
\end{equation}
gives the nature of the orbit to which the transformation 
converges starting with the initial perturbation (\ref{pert2}) 
on the critical surface. 
For instance, if we start with $\u{\theta}=\pi\u{\nu}_1$ 
which corresponds to the
perturbation (\ref{pert1}) with $M,P\leq 0$, the 
transformation converges to a cycle of period three
because $N^6\u{\theta}=\u{\theta} \mbox{ mod } 2\pi$. More precisely
this 3-cycle is the periodic sequence 
\begin{equation}
\{H_*(\u{\phi}+\pi\u{\nu}_1),H_*(\u{\phi}+\pi(\u{\nu}_1-\u{\nu}_2)),
H_*(\u{\phi}+\pi\u{\nu}_2)\}.
\end{equation}
For any orbit of the map (\ref{eqn:cat}), there is a fixed set that plays
the same role as the nontrivial fixed point for the even perturbation (\ref{pert1}). These
invariant sets belong to the same universality class as the fixed point, 
and in particular, they have the same critical exponents.\\
We can define a modified renormalization transformation such that the 
KAM-RG transformation converges to the fixed point $H_*$ for all initial 
perturbations of the form (\ref{pert2}) on the critical surface,
by adding an initial shift of the angles $\mathcal{T}_{-\theta}$.

\section{Conclusion}
\label{sect:conclusion}

The results show that the KAM-RG transformation is a powerful tool to describe
the breakup of invariant tori. In particular, the transformation describes 
with high accuracy the critical surface which is the stable manifold of a 
nontrivial fixed point (or more generally, of a nontrivial fixed set related
to this nontrivial fixed point by symmetries).\\
We have implemented the KAM-RG transformation for the torus with frequency
vector $\u{\omega}_0=(1/\goldenmean,-1)$. The extension to other 
frequencies that are quadratic irrationals is relatively clear. The case of
a general irrational frequency will involve qualitatively new features.
The KAM-RG transformation we described was for systems with two degrees of
freedom, but the extension to three~\cite{mackaymeissstark} or higher 
dimensional systems~\cite{koch} should be accessible.

\section*{acknowledgments}
We thank G. Gallavotti and G. Benfatto who initiated this approach in 1987 and
send us a preliminary report~\cite{gallbenf}. Our discussions with them
were very helpful for this project.
We also thank A. Celletti for providing the program 
to determine the critical coupling by Greene's criterion. We acknowledge
useful discussions with R. S. MacKay and H. Koch.
Support from the EC contract ERBCHRXCT94-0460 for the project
``Stability and universality in classical mechanics''
and from  the Conseil R\'egional de Bourgogne is acknowledged.


\begin{figure}
\large
\unitlength=1cm
\centerline{
\begin{picture}(15,10)
\put(1.5,0){\psfig{figure=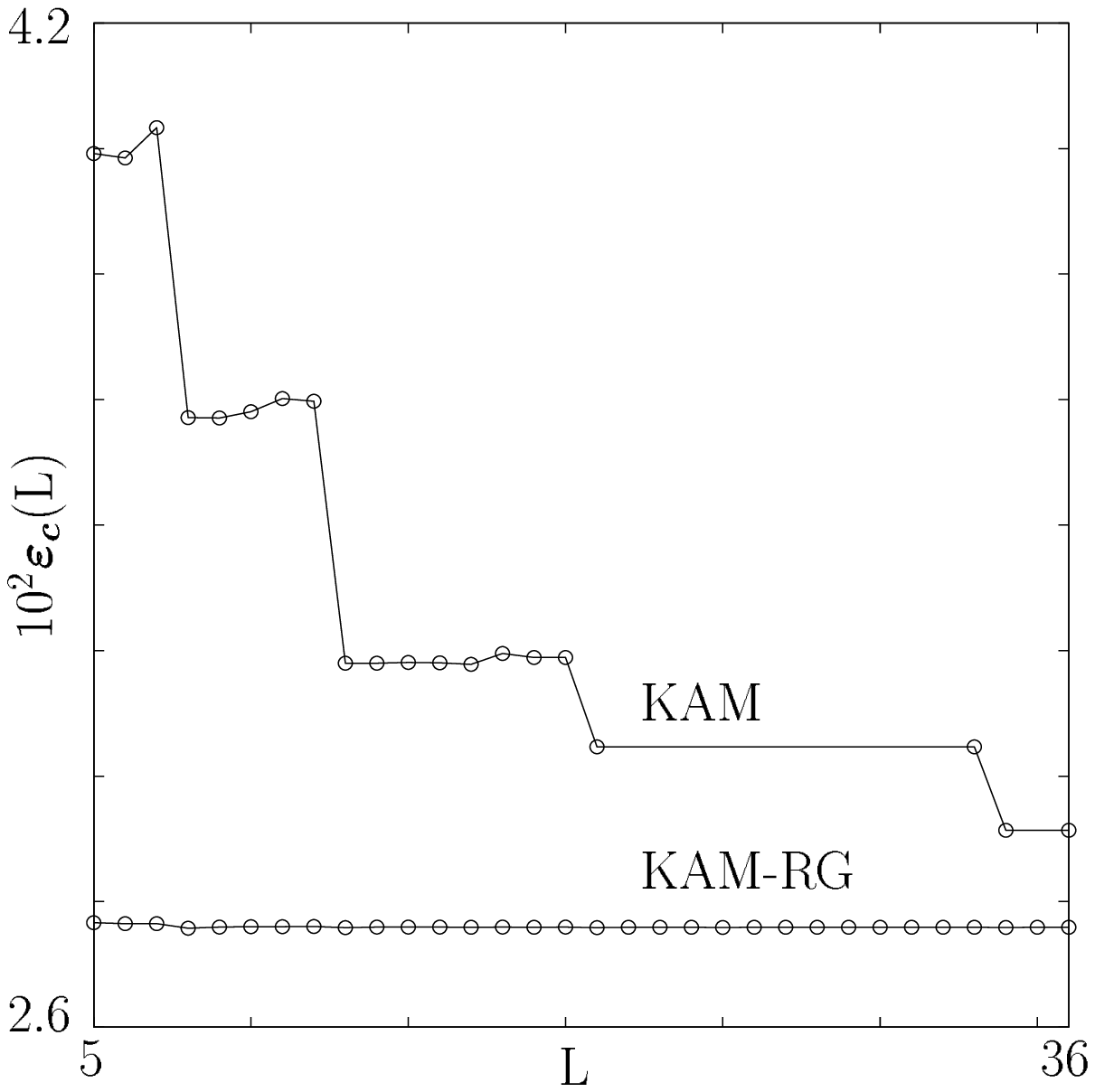,height=15cm,width=10cm}}
\normalsize 
\put(3,2){FIG.\ 1. Critical coupling $\varepsilon_c(L)$ as a function of $L$, the}
\put(3,1.6){size of the cell $\mathcal{C}_L$ containing $(2L+1)^2$ 
Fourier coefficients.}
\put(3,1.2){The upper curve corresponds to the KAM transformation,}
\put(3,0.8){and the lower one to the KAM-RG transformation.}
\end{picture}}
\end{figure}

\begin{figure} 
\large
\unitlength=1cm
\centerline{
\begin{picture}(15,10)
\put(1.5,0){\psfig{figure=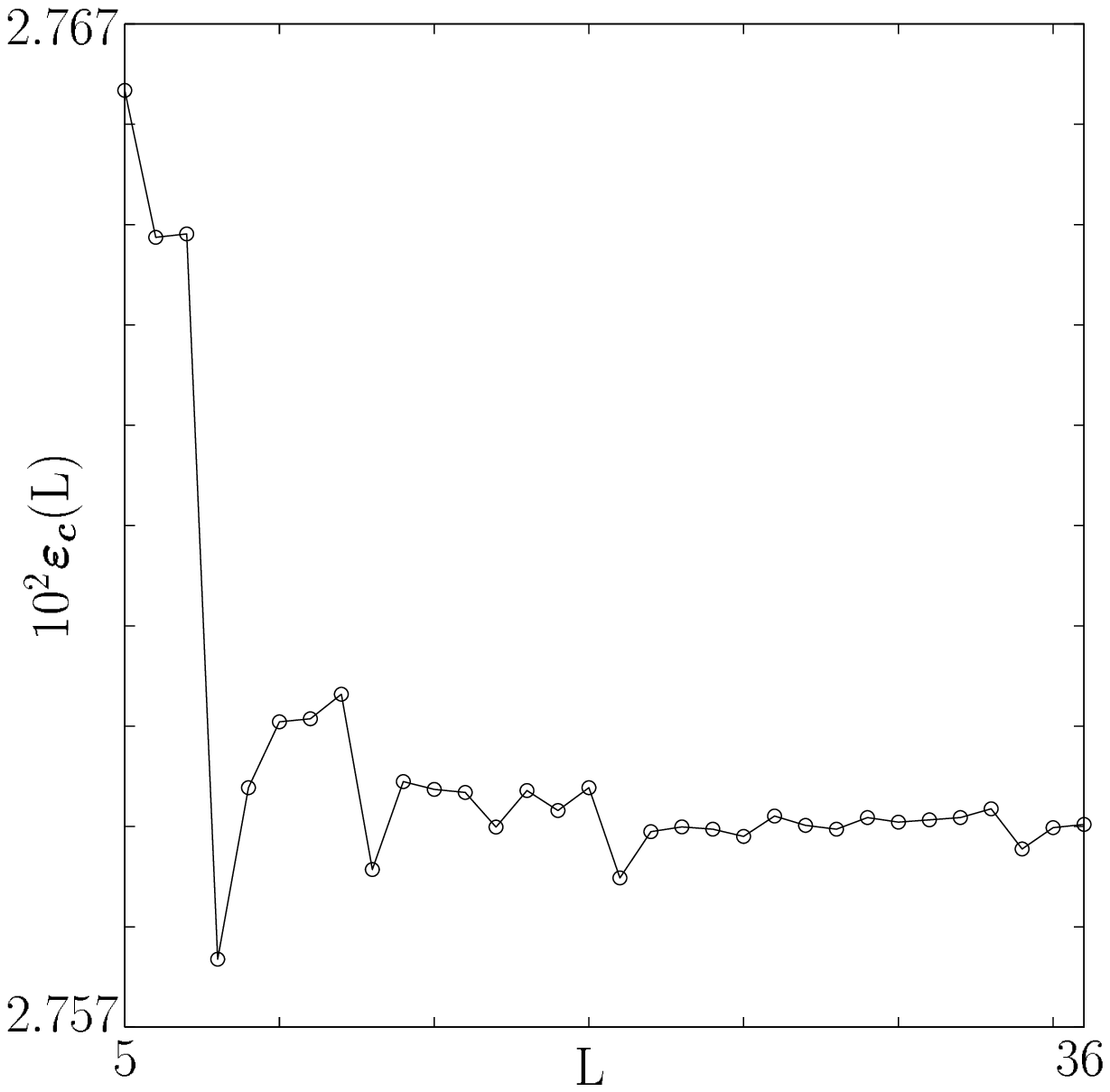,height=15cm,width=10cm}}
\normalsize 
\put(3,2){FIG.\ 2. Critical coupling $\varepsilon_c(L)$  of the KAM-RG}
\put(3,1.6){transformation as a function of $L$, the size of the }
\put(3,1.2){cell $\mathcal{C}_L$ containing $(2L+1)^2$ Fourier coefficients.}
\end{picture}}
\end{figure}

\begin{figure}
\large
\unitlength=1cm
\centerline{
\begin{picture}(15,10)
\put(1.5,0){\psfig{figure=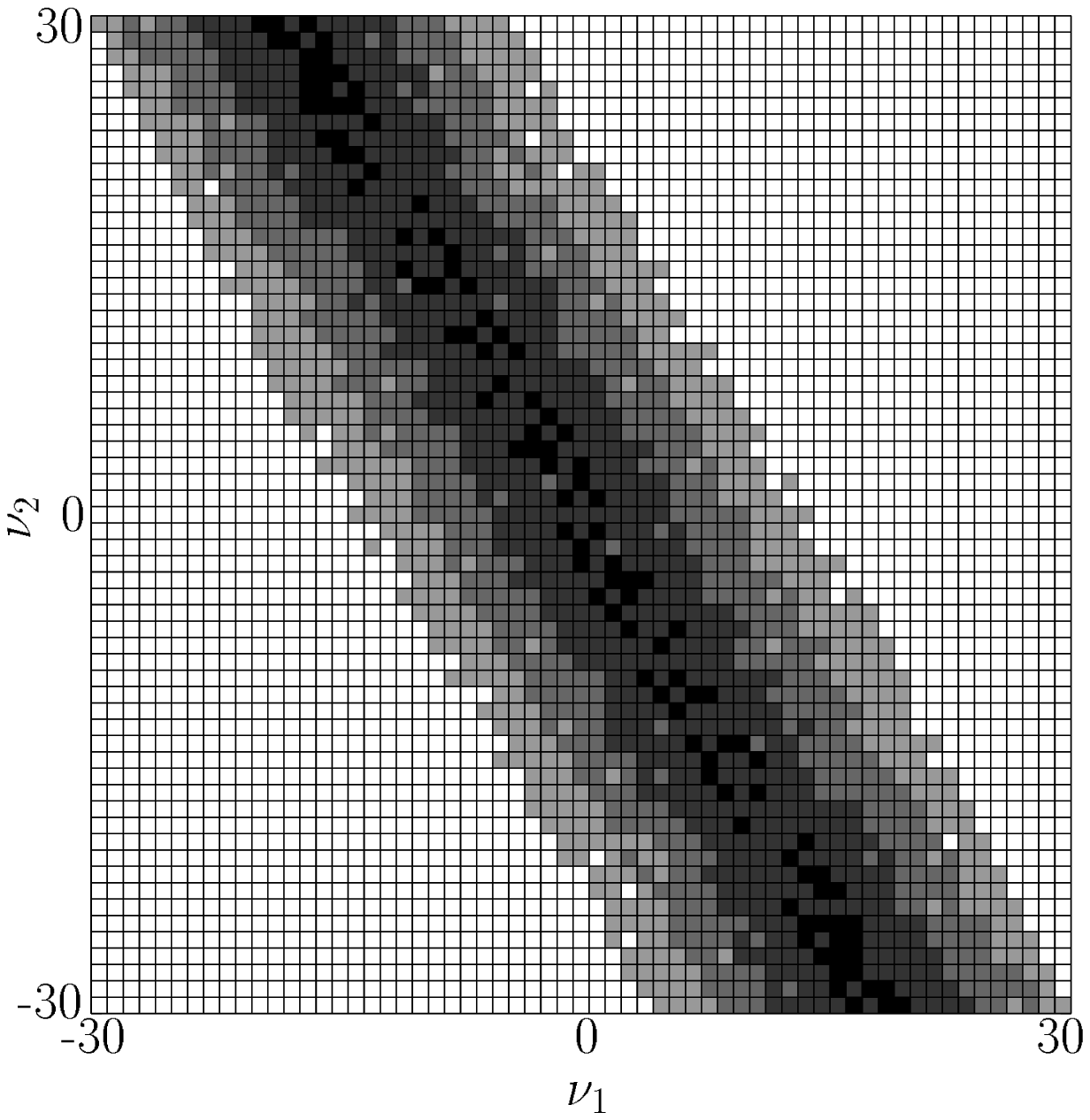,height=15cm,width=10cm}}
\normalsize 
\put(3,2){
FIG.\ 3. Weight of the Fourier coefficients of $f_*$:}
\put(3,1.6){White: $\, <10^{-10}$, grey levels: $[10^{-10},10^{-7}]$,}
\put(3,1.2){$[10^{-7},10^{-5}]$,$[10^{-5},10^{-3}]$, black: $[10^{-3},10^{-2}]$.}
\end{picture}}
\end{figure}

\begin{figure}
\large
\unitlength=1cm
\centerline{
\begin{picture}(15,10)
\put(1.5,0){\psfig{figure=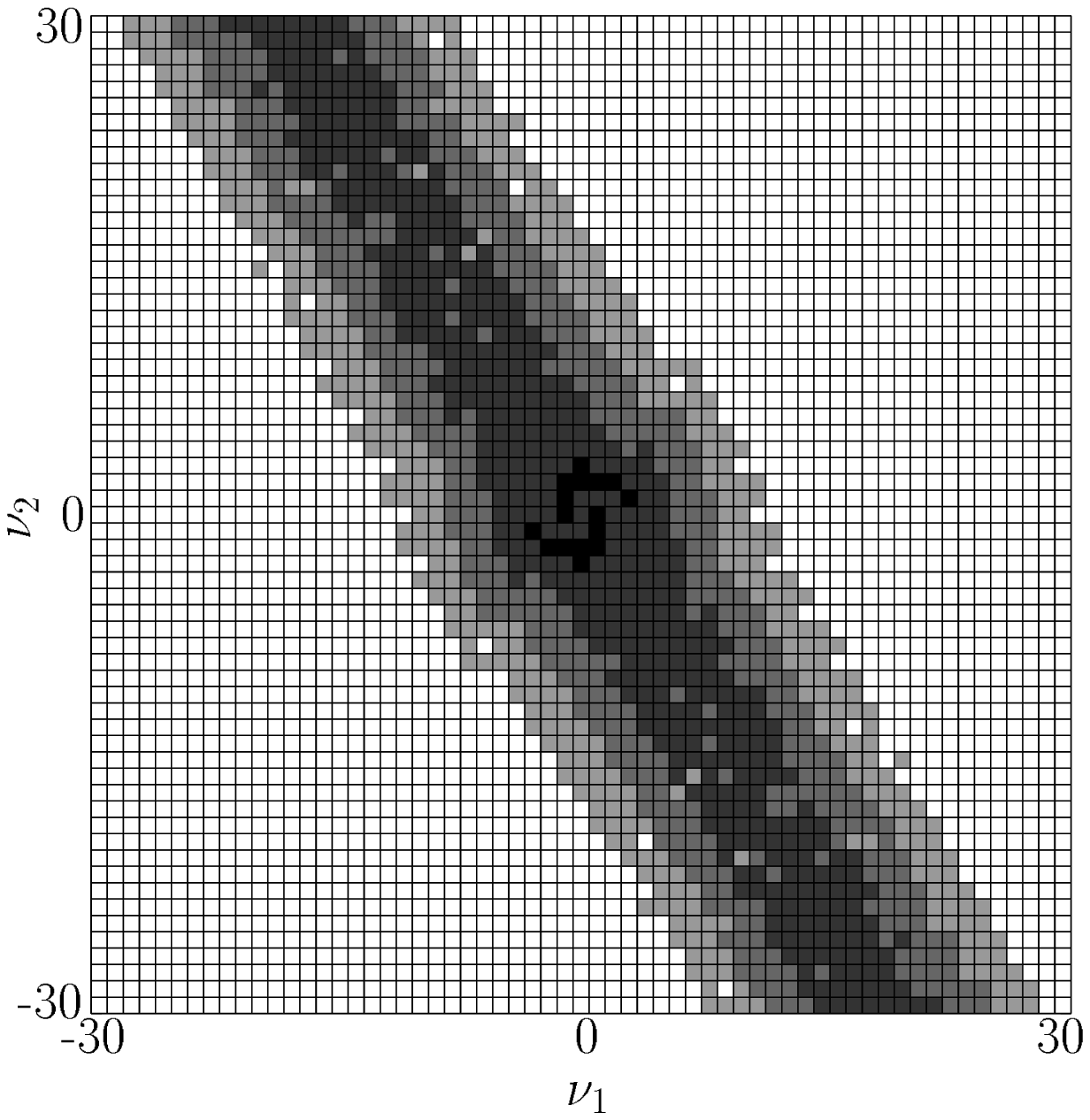,height=15cm,width=10cm}}
\normalsize 
\put(3,2){
FIG.\ 4. Weight of the Fourier coefficients of $g_*$:}
\put(3,1.6){White: $\, <10^{-10}$, grey levels: $[10^{-10},10^{-7}]$,}
\put(3,1.2){$[10^{-7},10^{-5}]$,$[10^{-5},10^{-3}]$, black: $[10^{-3},10^{-2}]$.}
\end{picture}}
\end{figure}

\begin{figure}
\large
\unitlength=1cm
\centerline{
\begin{picture}(15,10)
\put(1.5,0){\psfig{figure=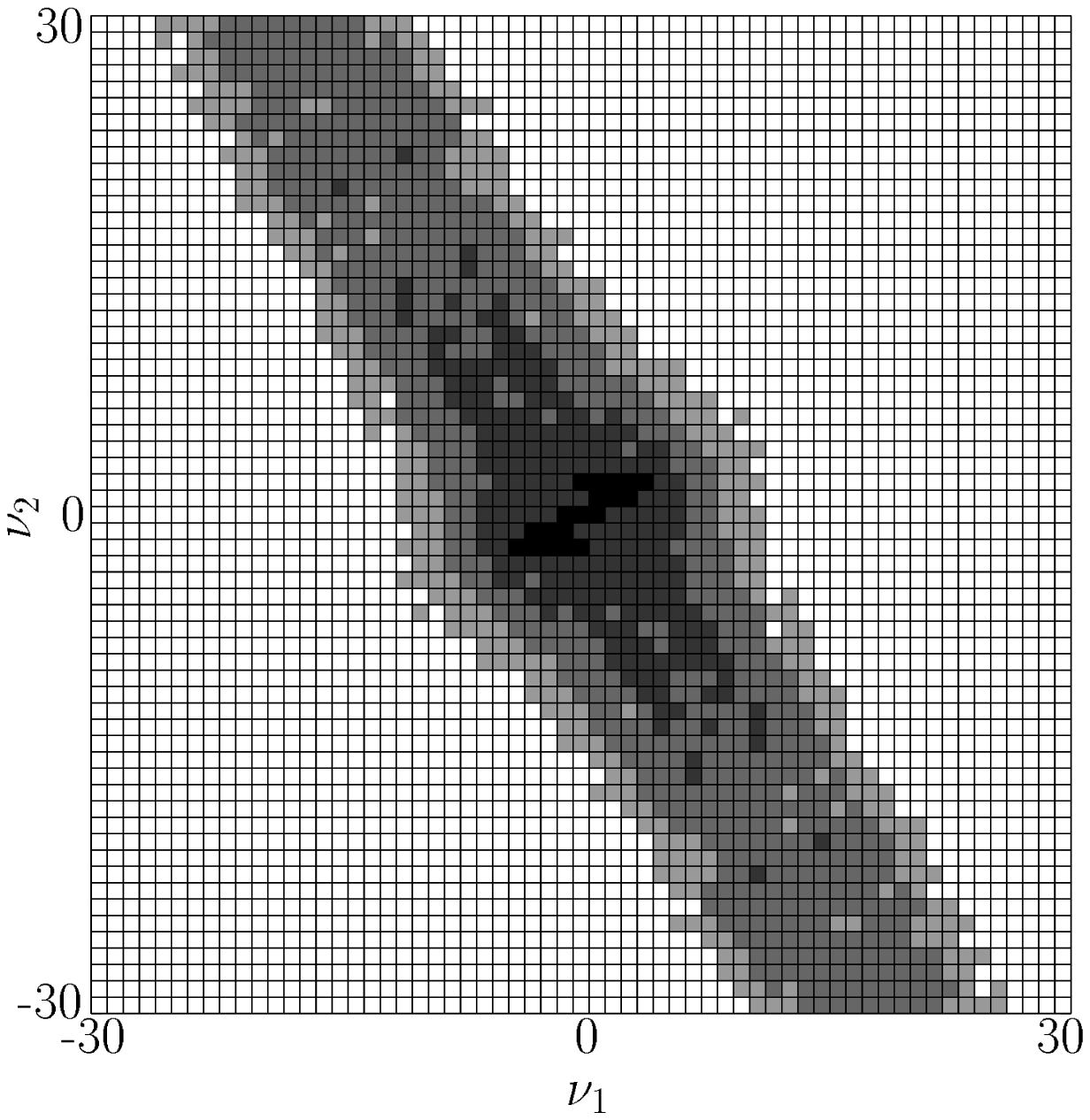,height=15cm,width=10cm}}
\normalsize 
\put(3,2){
FIG.\ 5. Weight of the Fourier coefficients of $m_*$:}
\put(3,1.6){White: $\, <10^{-10}$, grey levels: $[10^{-10},10^{-7}]$,}
\put(3,1.2){$[10^{-7},10^{-5}]$,$[10^{-5},10^{-3}]$, black: $[10^{-3},10^{-2}]$.}
\end{picture}}
\end{figure}



\newpage

\begin{references}

\bibitem{kolmogorov}
A.~N. Kolmogorov, Dokl. Akad. Nauk SSSR {\bf 98}, 527 (1954)
[in {\em Stochastic behaviour in classical and quantum
Hamiltonian systems}, edited by G. Casati and J. Ford,
Lecture Notes in Physics Vol. 93 (Springer,Berlin,1979), p. 51].

\bibitem{arnold}
V.~I. Arnold, Russ. Math. Surv. {\bf 18}, 9 (1963).

\bibitem{moser}
J.~Moser, {Nachr. Akad. Wiss. G\"ottinger}, 
Math.-Phys. Kl. IIa {\bf 1}, 1 (1962).

\bibitem{mackaypercival}
R.~S. MacKay and I.~C. Percival, Commun. Math. Phys. {\bf 98}, 469 (1985).

\bibitem{greene}
J.~M. Greene, J. Math. Phys. {\bf 20}, 1183 (1979).

\bibitem{falcolini}
C. Falcolini and R. de la Llave, J. Stat. Phys. {\bf 67}, 609 (1992).

\bibitem{mackay3}
R. S. MacKay, Nonlinearity {\bf 5}, 161 (1992).

\bibitem{aubry}
S.~Aubry and P.~Y. Le~Daeron, Physica D {\bf 8}, 381 (1983).

\bibitem{mather}
J.~N. Mather, Topology {\bf 21}, 457 (1982).

\bibitem{moser2}
J. Moser, SIAM Review {\bf 28}, 459 (1986).

\bibitem{denzler}
J. Denzler, J. Appl. Math. and Phys. {\bf 38}, 791 (1987).

\bibitem{kadanoff}
L.~P. Kadanoff, Phys. Rev. Lett. {\bf 47}, 1641 (1981).

\bibitem{shenkerkadanoff}
S.~J. Shenker and L.~P. Kadanoff, J. Stat. Phys. {\bf 27}, 631 (1982).

\bibitem{rand}
D.~Rand, S.~Ostlund, J.~Sethna, and E.~D. Siggia, Phys. Rev. Lett. {\bf 49},
132 (1982).

\bibitem{mackay}
R.~S. MacKay, Physica D {\bf 7}, 283 (1983).

\bibitem{mackayL}
R. S. MacKay, {\em Renormalisation in Area-preserving Maps} (World Scientific,
1993).

\bibitem{escandedoveil}
D.~F. Escande and F.~Doveil, J. Stat. Phys. {\bf 26}, 257 (1981).

\bibitem{escande}
D.~F. Escande, Phys. Rep. {\bf 121}, 165 (1985).

\bibitem{koch}
H.~Koch, archived in mp\_arc@math.utexas.edu, \#96-383 (1996).

\bibitem{govin}
M.~Govin, C.~Chandre, and H.~R.~Jauslin, Phys. Rev. Lett. {\bf 79}, 3881 (1997).

\bibitem{thirring}
W.~Thirring, {\em A Course in Mathematical Physics I: Classical Dynamical
  Systems} (Springer Verlag, Berlin, 1992), p. 153.

\bibitem{gallavotti2}
G.~Gallavotti, Commun. Math. Phys. {\bf 164}, 145 (1994).

\bibitem{deprit}
A.~Deprit, Celest. Mech. {\bf 1}, 12 (1969).

\bibitem{benettin}
G. Benettin, L. Galgani, A. Giorgilli, and J. M. Strelcyn, Nuovo Cimento
{\bf 79}B, 201 (1984).

\bibitem{goldstein}
H. Goldstein, {\em Classical Mechanics} (Addison-Wesley, Reading, Mass., 1980).

\bibitem{cellettichierchia}
A.~Celletti and L.~Chierchia, Commun. Math. Phys. {\bf 118}, 119 (1988).

\bibitem{escandeetal}
D.~F. Escande, M.~S. Mohamed-Benkadda, and F.~Doveil, Phys. Lett. A
{\bf 101}, 309 (1984).

\bibitem{greenemao}
J.~M. Greene and J.~Mao, Nonlinearity {\bf 3}, 69 (1990).

\bibitem{mackay2}
R.~S. MacKay, in {\em Proceedings of the International Conference on
Dynamical Systems and Chaos, Tokyo}, 
edited by Y. Aizawa, S. Saito and K. Shiraiwa (World Scientific, 1995), p. 34.

\bibitem{arnoldavez}
V. I. Arnold and A. Avez, {\it Ergodic Problems of Classical Mechanics}
(Benjamin, New-York, 1968).

\bibitem{mackaymeissstark}
R.~S. MacKay, J.~D. Meiss, and J.~Stark, Phys. Lett. A {\bf 190}, 417 (1994).

\bibitem{gallbenf}
G. Gallavotti and G. Benfatto (unpublished).

\end{references}
\end{document}